\documentclass[aps,prl,twocolumn,preprintnumbers,amsmath,amssymb,superscriptaddress]{revtex4}%

\usepackage{graphicx}%
\usepackage{dcolumn}
\usepackage{amsmath}
\usepackage{color}
\usepackage{multirow}

\begin{document}



\title{Pressure-induced electronic phase separation of magnetism and superconductivity in CrAs}

\author{R.~Khasanov}
 \email[Corresponding author: ]{rustem.khasanov@psi.ch}
 \affiliation{Laboratory for Muon Spin Spectroscopy, Paul Scherrer Institute, CH-5232 Villigen PSI, Switzerland}
\author{Z.~Guguchia}
 \affiliation{Laboratory for Muon Spin Spectroscopy, Paul Scherrer Institute, CH-5232 Villigen PSI, Switzerland}
\author{I.~Eremin}
 \affiliation{Institut f\"{u}r Theoretische Physik III, Ruhr-Universit\"{a}t Bochum, D-44801 Bochum, Germany}
\author{H.~Luetkens}
 \affiliation{Laboratory for Muon Spin Spectroscopy, Paul Scherrer Institute, CH-5232 Villigen PSI, Switzerland}
\author{A.~Amato}
 \affiliation{Laboratory for Muon Spin Spectroscopy, Paul Scherrer Institute, CH-5232 Villigen PSI, Switzerland}
\author{P.K.~Biswas}
 \affiliation{Laboratory for Muon Spin Spectroscopy, Paul Scherrer Institute, CH-5232 Villigen PSI, Switzerland}
\author{Ch.~R\"{u}egg}
 \affiliation{Laboratory for Neutron Scattering and Imaging, Paul Scherrer Institute, CH-5232 Villigen PSI, Switzerland}
 \affiliation{Department of Condensed Matter Physics, University of Geneva, CH-1211 Geneva, Switzerland}
\author{M.A.~Susner}
 \affiliation{Materials Science and Technology Division, Oak Ridge National Laboratory, Oak Ridge, TN 37831-6114, USA}
\author{A.S.~Sefat}
 \affiliation{Materials Science and Technology Division, Oak Ridge National Laboratory, Oak Ridge, TN 37831-6114, USA}
\author{N.D.~Zhigadlo}
 \affiliation{Solid State Physics Laboratory, ETH Zurich, 8093 Zurich, Switzerland}
\author{E.~Morenzoni}
 \affiliation{Laboratory for Muon Spin Spectroscopy, Paul Scherrer Institute, CH-5232 Villigen PSI, Switzerland}

\begin{abstract}
The recent discovery of pressure induced superconductivity in the binary helimagnet CrAs has attracted much attention. How superconductivity emerges from the magnetic state and what is the mechanism of the superconducting pairing are two important issues which need to be resolved. In the present work, the suppression of magnetism and the occurrence of superconductivity in CrAs as a function of pressure ($p$) were studied by means of muon spin rotation. The magnetism remains bulk up to $p\simeq3.5$~kbar while its volume fraction gradually decreases with increasing pressure until it vanishes at $p\simeq$7~kbar. At 3.5 kbar superconductivity abruptly appears with its maximum $T_c \simeq 1.2$~K which decreases upon increasing the pressure. In the intermediate pressure region ($3.5\lesssim p\lesssim 7$~kbar) the superconducting and the magnetic volume fractions are spatially phase separated and compete for phase volume. Our results indicate that the less conductive magnetic phase provides additional carriers (doping) to the superconducting parts of the CrAs sample thus leading to an increase of the transition temperature ($T_c$) and of the superfluid density ($\rho_s$).
A scaling of $\rho_s$ with $T_c^{3.2}$ as well as the phase separation between magnetism and superconductivity point to a
conventional mechanism of the Cooper-pairing in CrAs.
\end{abstract}
\maketitle

The pressure-induced superconductivity in the binary helimagnet CrAs has recently attracted much attention
\cite{Wu_Arxiv_14,Kotegawa_JPSJ_14,Kotegawa_Arxiv_14,Keller_Arxiv_14,Shen_Arxiv_14}. At ambient pressure CrAs is
characterized by a relatively high N\'{e}el temperature $T_N\simeq 270$~K \cite{Watanabe_69,Selte_71,Boller_71}.
$T_N$  decreases approximately  by a factor of three for pressures ($p$) approaching $\simeq7$~kbar, above which the magnetism
completely disappears \cite{Wu_Arxiv_14,Kotegawa_JPSJ_14,Kotegawa_Arxiv_14}. On the other hand superconductivity sets in for
pressures exceeding $\simeq4$~kbar thus revealing a range of $4\lesssim p \lesssim 7$~kbar where superconductivity
and magnetism coexist.

The close proximity of superconductivity to magnetism, the similarity of the phase diagram of CrAs with that of some Fe-based
superconductors, as well as the absence of the coherent Hebel-Slichter peak in the nuclear relaxation rate $1/T_1T$ made the
authors of Refs.~\onlinecite{Wu_Arxiv_14,Kotegawa_JPSJ_14,Kotegawa_Arxiv_14,Shen_Arxiv_14} to suggest an unconventional
pairing mechanism.
It should be noted, however, that the similarity of the phase diagram does not necessarily requires a similar mechanism of
Cooper-pairing. The Hebel-Slichter peak can also be suppressed in conventional $s-$wave superconductors. This is {\it e.g.}
the case for superconductors in the strong coupling limit \cite{Ohsugi_JPSJ_91}, for superconductors having  a spread of $T_c$ over the sample, or for slight gap anisotropies \cite{Stenger_PRL_91}. Whether a coherence peak is present at all in the archetypical two-gap
superconductor MgB$_2$ is still subject of discussion \cite{Jung_PRB_01,Baek_PRB_02}. %
One needs, therefore, a more detailed investigation of the superconducting response of CrAs as well as an understanding on
how superconductivity emerges from a compound being initially in a strong magnetic state.

In this paper we report on muon spin rotation ($\mu$SR) studies of the magnetic and the superconducting properties of CrAs.
We first discuss separately the magnetic and the superconducting responses as a function of pressure, and concentrate later
on the issue of coexistence between magnetism and superconductivity.

\vspace{0.2cm}
\noindent{\bf Results }\\
\noindent{\it Magnetism in CrAs. }
The magnetic response of CrAs powder samples was studied by zero field (ZF) and weak transverse field (wTF) $\mu$SR
experiments. In the following we discuss the $\mu$SR data for three different pressure regions.

\begin{figure}[t]
\includegraphics[width=0.95\linewidth]{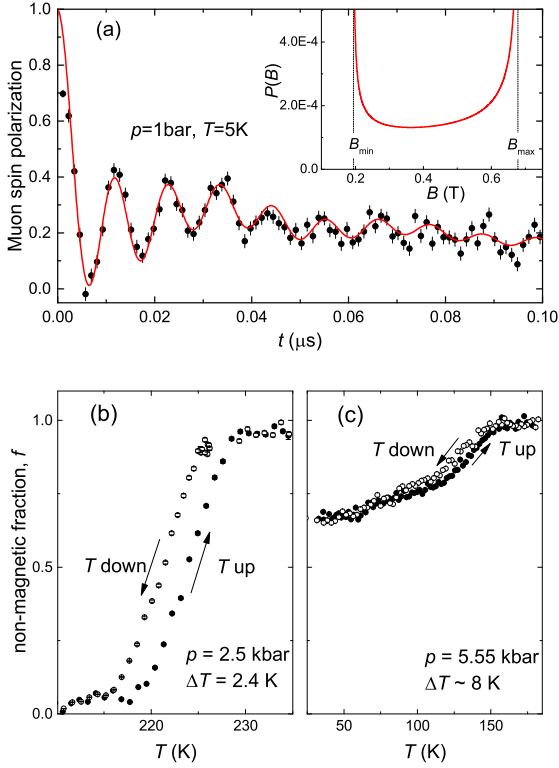}
 \vspace{-.5cm}
\caption{{\bf Representative ZF and wTF $\mu$SR data.} {\bf(a)} ZF $\mu$SR time-spectra of CrAs measured at $T=5$~K and $p=1$~bar. The
solid line is a fit according to the theoretical field distribution caused by incommensurate helimagnetic order shown in the
inset [see the Supplementary materials for details]. The minimum ($B_{min}$) and the maximum($B_{max}$) cutoff fields are
represented by vertical dashed lines. {\bf (b)} and {\bf (c)} depict the temperature evolution of the non-magnetic volume fraction $f$
of CrAs obtained in the wTF $\mu$SR measurements at $p=2.5$ and 5.55~bar, respectively. Closed and open symbols correspond to
the experimental data obtained with increasing and decreasing the temperature. The clear hysteresis is indicative of a first order
magnetic transition. }
 \label{fig:ZF-WTF}
\end{figure}

In the low-pressure region, $1$~bar$\leq p\lesssim3.5$~kbar, spontaneous muon spin precession is clearly seen in the ZF
$\mu$SR time spectra (see Fig.~\ref{fig:ZF-WTF}~a) thus confirming that long range magnetic order is established below $T_N$.
The oscillating part of the signal is accurately described by a field distribution characterized by a minimum ($B_{min}$) and
a maximum ($B_{max}$) cutoff field (see the inset in Fig.~\ref{fig:ZF-WTF}~a), which is consistent with the observation of
helimagnetic incommensurate magnetic order \cite{Keller_Arxiv_14, Shen_Arxiv_14, Watanabe_69, Selte_71, Boller_71}.
The relatively high values of the cutoff fields ($B_{min}\simeq0.194$~T and $B_{max}\simeq0.678$~T at $p=1$~bar) are in agreement with the large moments $m(1$~bar$)\simeq 1.73$~$\mu_B$ as obtained by means of  neutron powder
diffraction \cite{Keller_Arxiv_14}. The wTF $\mu$SR experiments performed at ambient pressure and at $p=2.5$~kbar show
relatively sharp transitions to the magnetic state and prove that the magnetism occupies close to 100\% of the sample volume (see
Fig.~\ref{fig:ZF-WTF}~b  and Fig.~Sup~3 in the Supplementary material). The hysteresis in $T_N$ signifies a first order magnetic phase transition.

In the intermediate pressure region ($3.5\lesssim p\lesssim7$~kbar) the cutoff fields, which are proportional to the ordered moment,  decrease continuously and reach at
$p\simeq6.7$~kbar $B_{min}\simeq0.14$~T and $B_{max}\simeq0.58$~T  (see Fig.~Sup~1 in the Supplementary materials). This is
consistent with a decrease of the ordered magnetic moment to $m(6.7$~kbar$)\simeq 1.47$~$\mu_B$. The wTF $\mu$SR
experiments reveal that the low temperature value of the non-magnetic fraction $f$ gradually increases with increasing pressure by
reaching $f\simeq 1$ for pressures exceeding 7~kbar (see Fig.~\ref{fig:ZF-WTF}~c and Fig.~Sup~3 in the Supplementary materials).
Therefore in the intermediate pressure region the sample is  separated into a magnetically ordered phase and a
non-magnetic phase. The hysteresis in $T_N$ confirms that the magnetic transition remains of first order at all pressures (see
Fig.~\ref{fig:ZF-WTF}~c).

For pressures above  7~kbar the ZF $\mu$SR experiments prove the absence of any type of magnetic order as exemplified by the
weakly damped wTF $\mu$SR time spectra.

\vspace{0.1cm}
\noindent{\it Superconductivity in CrAs. }
\begin{figure}[t]
\includegraphics[width=0.85\linewidth]{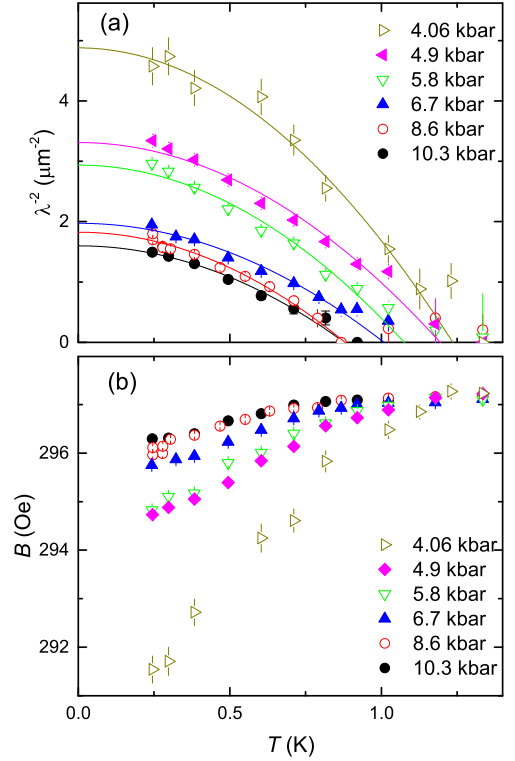}
 \vspace{-0.5cm}
\caption{{\bf The superfluid density and the diamagnetic shift at various pressures.} Temperature evolution of the inverse squared magnetic penetration depth $\lambda^{-2}\propto\rho_s$ {\bf (a)} and the internal
field $B$ {\bf (b)} obtained from the fit of 30~mT TF $\mu$SR data measured at $p=4.06$, 4.9, 5.8, 6.7, 8.6, and 10.3~kbar.
Solid lines in (a) are power law fits $\lambda^{-2}(T)=\lambda^{-2}(0)[1-(T/T_c)^n]$ with a common exponent $n=1.95(3)$ for
all data sets. }
 \label{fig:superfluid}
\end{figure}
The superconducting response of CrAs was studied in transverse field (TF) $\mu$SR experiments (applied field $\mu_0H=30$~mT).
From the experimental data we have extracted the magnetic penetration depth $\lambda$, which is  related to the superfluid
density $\rho_s$ in terms of $\rho_s = n_s/m^\ast\propto\lambda^{-2}$ ($n _s$ charge carrier concentration and $m^\ast$
carrier effective mass).
The magnetic penetration depth $\lambda$ was determined from the Gaussian muon-spin depolarization rate $\sigma_{sc}(T)
\propto \lambda^{-2}(T)$, which reflects the second moment of the magnetic field distribution in the superconductor in the
mixed state \cite{Zimmermann_PRB_95}. $\sigma_{sc}$ is related to $\lambda$ via
$\sigma_{sc}^2/\gamma_\mu^2=0.00371\Phi_0^2\lambda^{-4}$ \cite{Brandt_PRB_88}
($\Phi_0=2.068\cdot10^{-15}$~Wb is the magnetic flux quantum, and $\gamma_{\mu}= 2\pi\cdot135.5$~MHz/T is the muon
gyromagnetic ratio).

\begin{figure}[t]
\includegraphics[width=1.0\linewidth]{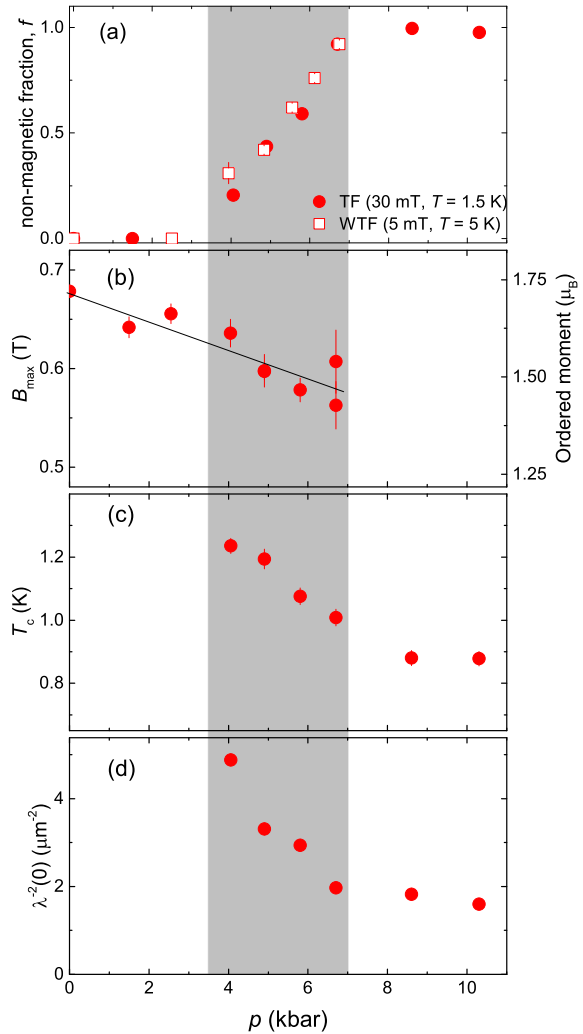}
 \vspace{-0.8cm}
\caption{{\bf Temperature-pressure phase diagram.} Pressure dependence of the non-magnetic volume fraction $f$ {\bf (a)}; maximum cutoff field $B_{max}$, which is proportional to the ordered moment $B_{max}\propto m$  {\bf (b)};
superconducting transition temperature $T_c$ {\bf (c)}; and the zero-temperature value of the inverse squared magnetic
penetration depth $\lambda^{-2}(0)\propto\rho_s$ {\bf (d)}. The grey area represents the pressure region where magnetism and
superconductivity coexist. The solid line in {\bf (b)} is a linear fit with $B_{max}(p)=0.6782(14)-0.0153(5)p$ (see the
Supplementary materials). }
 \label{fig:results}
\end{figure}

The measured $\lambda^{-2}(T)$ and the internal field $B(T)$ of CrAs for $p=4.06$, 4.9, 5.8, 6.7, 8.6 and 10.3~kbar are shown
in Figs.~\ref{fig:superfluid}~a and \ref{fig:superfluid}~b. Note that $\lambda^{-2}$ and $B$ were derived from the fraction
of the sample remaining in the non-magnetic state down to the lowest temperature (see Fig.~\ref{fig:results}~a). Due to the
strongly damped signal in the magnetic phase one is unable to measure any superconducting response in the magnetic fraction
of the sample. We believe, however that superconductivity in CrAs cannot emerge in the magnetically ordered parts for two
following reasons. First, Wu {\it et al.} \cite{Wu_Arxiv_14} have shown that the low-temperature diamagnetic susceptibility
($\chi_{dia}$) of CrAs is nearly zero for pressures $p\lesssim 3$~kbar, increases linearly in the range $4.09\lesssim
p\lesssim 7.85$ and reaches its maximum value, close to the ideal $\chi_{dia}=-1/4\pi$, for pressures exceeding 7.85~kbar. It
follows almost exactly the pressure dependence of the non-magnetic fraction $f$ as observed in our wTF and TF $\mu$SR
experiments (see Fig.\ref{fig:results}~a) \cite{comment1}. Second, the large magnetic moment and its weak reduction as a
function of pressure (see Fig.~\ref{fig:results}~b) require the separation of CrAs in superconducting and magnetic domains.
This is {\it e.g.} the case for the so-called '245' family of Fe-based superconductors \cite{Li_NatPh_12,Maletz_PRB_13},
which is characterized by the high value of both, magnetic moment ($\sim 3$~$\mu_B$) and N\'{e}el temperature
($T_N\sim500$~K) \cite{Shermadini_PRL_11,Bao_ChPL_11,Wang_NatCom_11,Li_PRB_11}. Note that within the full pressure range
studied here the value of the ordered magnetic moment in CrAs is only a factor of two smaller than that in '245'
superconductors.

The absence of experimental points below $T\simeq0.24$~K does not allow us to draw any conclusion about the possible gap
symmetry in CrAs based on the $\lambda^{-2}(T)$ data. Therefore, they were fitted to a power law
$\lambda^{-2}(T)=\lambda^{-2}(0)[1-(T/T_c)^n]$ with the common exponent $n=1.95(3)$ for all data sets. Values for the
superconducting transition temperature $T_c$ and the inverse squared zero-temperature magnetic penetration depth
$\lambda^{-2}(0)$ obtained from these fits are plotted in Figs.~\ref{fig:results}~c and ~\ref{fig:results}~d.

\vspace{0.2cm}
\noindent{\bf Discussion}\\
\noindent{\it Interplay between magnetism and superconductivity.}
Figure~\ref{fig:results} summarizes our results on the magnetism and superconductivity in CrAs as a function of pressure.
CrAs remains purely magnetic up to $p\simeq3.5$~kbar. Above this pressure and up to $p\simeq7$~kbar both, magnetic and
superconducting responses are clearly detected in a set of ZF, wTF, and TF $\mu$SR experiments. CrAs is phase separated into
volumes where long range magnetic order is established below the N\'{e}el temperature $T_N$ and into non-magnetic volumes becoming superconducting below the critical temperature $T_c$. 
It is interesting to note that, besides the competition for the volume, there is no evidence for a competition between the magnetic and superconducting order parameter in CrAs. This is in
contrast to various Fe-based and cuprate superconductors where it is generally observed. Indeed, the ordered magnetic moment
stays almost {\it constant}, by changing less than 15\% from 1.73~$\mu_B$ at $p=1$~bar to 1.47~$\mu_B$ at $p\simeq 6.7$~kbar, see
Fig.~\ref{fig:results}~b. $T_N$, in their turn, evolves smoothly with pressure without showing any pronounced features at $p\simeq 3.5-4$~kbar, {\it i.e.} where the non-magnetic phase starts to develop (see
Refs.~\onlinecite{Wu_Arxiv_14,Kotegawa_JPSJ_14,Kotegawa_Arxiv_14,Shen_Arxiv_14}).

Figs.~\ref{fig:results}~a and \ref{fig:results}~d demonstrate that the maximum value of the superfluid density
$\rho_s\propto\lambda^{-2}$ is observed at the low pressure side of the phase separated region {\it i.e.} in the region where
the non-magnetic volume fraction $f$ is the smallest. With further increasing $f$, the superfluid density decreases until it
saturates when $f\simeq 1$.   By neglecting the pressure effect on the charge carrier mass $m^\ast$, the superfluid density is
simply proportional to the carrier concentration $\rho_s\propto n_s$. We may assume, therefore, that within the phase
separated region carriers from the 'less conductive' magnetically ordered parts of the sample can be supplied to the 'more
conductive' non-magnetic parts, which become superconducting at low temperatures.
The effect of supplying additional carriers, which can be considered as "doping", is expected to be the strongest if the
magnetic volume fraction exceeds substantially the paramagnetic one ($f\ll 1$), while it should decrease and even vanishes
completely for $f$ approaching 1. Figs.~\ref{fig:results}~a and \ref{fig:results}~d imply that this is exactly the case for
CrAs. Effectively, the non-magnetic volume fraction $f$ {\it anticorrelates} with the superfluid density $\rho_s$.

\begin{figure}[t]
\includegraphics[width=.95\linewidth]{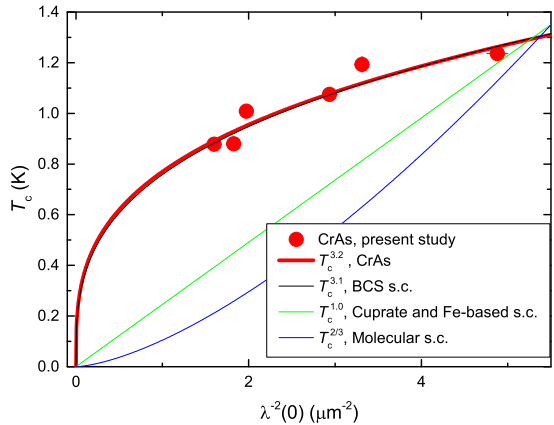}
 \vspace{-.5cm}
\caption{{\bf Correlation between $T_c$ and $\lambda^{-2}(0)$.} Superconducting critical temperature $T_c$ versus inverse squared zero-temperature magnetic penetration depth
$\lambda^{-2}(0)$ of CrAs. The red line is $\lambda^{-2}(0)\propto (T_c)^n$ fit to CrAs data with the exponent $n=3.2(2)$.
The black, green and blue lines are empirical relations for some phonon mediated BCS superconductors ($n=3.1$,
Ref.~\onlinecite{Khasanov_PRB_08_2}), cuprate and Fe-Based high-temperature superconductors ($n=1$,
Refs.~\onlinecite{Uemura_PRL_89,Uemura_PRL_91,Luetkens_PRL_08,Khasanov_PRB_08,Goko_PRB_09,Pratt_PRB_09}) and molecular
superconductors ($n=2/3$, Ref.~\onlinecite{Pratt_PRL_05}), respectively. }
 \label{fig:Tc-lambda_relation}
\end{figure}

\vspace{0.1cm}
\noindent{\it Correlation between $T_c$ and $\lambda^{-2}$.}
Figs.~\ref{fig:results}~c and \ref{fig:results}~d show that $T_c$ and $\lambda^{-2}(0)$ have similar pressure dependences,
which could point to a possible correlation between these quantities. The famous "Uemura line" establishes a linear relation
between $T_c$ and $\lambda^{-2}(0)$ for various families of underdoped cuprate high-temperature superconductors
\cite{Uemura_PRL_89,Uemura_PRL_91}. A similar linear relation was observed in recently discovered Fe-based superconductors
\cite{Luetkens_PRL_08,Khasanov_PRB_08,Goko_PRB_09,Pratt_PRB_09}. In molecular superconductors  $\lambda^{-2}(0)$ was found to
be proportional to $T_c^{2/3}$ \cite{Pratt_PRL_05}, while in some phonon mediated BCS superconductors $\lambda^{-2}(0)\propto
T_c^{3.1}$ \cite{Khasanov_PRB_08_2}. Figure \ref{fig:Tc-lambda_relation} shows that in CrAs $\lambda^{-2}(0)$ the data scales
as $T_c^{3.2(2)}$ thus suggesting that superconductivity in CrAs is most probably BCS like and is mediated by phonons.

A further indication of conventional electron-phonon coupling in CrAs comes from the observed macroscopic phase separation of
the magnetic and the superconducting phases.
Following Ref.~\onlinecite{Fernandes_PRB_10}, the relative phase difference ($\theta$) of the superconducting order parameter
between different parts of Fermi surface or Fermi surface sheets  may lead either to  stabilization of microscopic
coexistence of the magnetic and superconducting phases or drive both to repel each other. This happens because the staggered
magnetic moment ({\bf M}) plays in a superconductor the role of an intrinsic Josephson coupling with the free energy term
$F_J \propto {\bf M}^2 |\Delta_1||\Delta_2| \cos \theta$
($\Delta_1$, $\Delta_2$ are the superconducting order parameters at different parts of the Fermi surface or Fermi surface
sheets). If the superconducting order is a conventional one, ({\it i.e.} there is no internal phase change,  $\theta=0$),
this term increases the total energy, thus making both phases unlikely to coexist. On the contrary, if the phases are
opposite such that $\theta=\pi$ the Josephson coupling term in the free energy is negative. As a result both the
superconducting and the magnetic phases like to coexist. This explains why the magnetic and superconducting orders do coexist
microscopically in some unconventional superconductors  like ferropnictides, electron-doped cuprates, and heavy-fermion
systems where the order parameter has an internal phase shift. In CrAs, however, the phase diagram points towards an
isotropic $s-$wave symmetry of the superconducting order parameter driven by electron-phonon interaction.

\vspace{0.2cm}
\noindent{\bf Conclusions }\\
To conclude, the magnetic and the superconducting properties of CrAs as a function of pressure were studied by means of muon
spin rotation. The bulk magnetism exists up to $p\simeq3.5$~kbar, while the purely non-magnetic state develops for pressures
above $\simeq7$~kbar. In the intermediate pressure region ($3.5\lesssim p\lesssim 7$~kbar) the magnetic phase volume
decreases continuously and superconductivity develops in parts of the sample remaining non-magnetic down to the lowest
temperatures. Both, the superconducting transition temperature $T_c$ and the zero-temperature superfluid density $\rho_s(0)$
decrease with increasing pressure in the intermediate pressure region  and saturate for $p$ exceeding $\simeq7$~kbar {\it
i.e.} in the region where magnetism  is completely suppressed.

Our results suggest that the pressure-induced transition of CrAs from a magnetic to a superconducting state is characterized
by a separation in macroscopic size magnetic and superconducting volumes. The less conductive magnetic phase provides
additional carriers (doping) to the superconducting parts of CrAs. This would naturally explain the substantial increase of both, the transition temperature $T_c$ (from 0.9~K to 1.2~K) and the superfluid density $\rho_s(0)$ (up to $\simeq 150$\%), in the
phase coexistence region. The superfluid density was found to scale with $T_c$ as $\rho_s\propto T_c^{3.2(2)}$, which,
together with the clear phase separation between magnetism and superconductivity, points towards a conventional mechanism of
the Cooper-pairing in CrAs.

\newpage

\vspace{1cm}
\noindent {\bf Methods}\\
\noindent {\it Sample preparation.}
Two type of policristalline CrAs samples were used during our studies.
The first type of samples was prepared by means of high-pressure synthesis. Overall details of the sample cell assembly and high-pressure synthesis process can be found in Ref.~\onlinecite{Zhigadlo_PRB_12}. The mixture of Cr (99.9\%) and As (99.99\%) powders in a molar ratio 1:1 was enclosed in a boron nitride (BN) crucible and placed inside a pyrophylite cube with a graphite heater. In a typical run, the sample was compressed to 15~kbar at room temperature. While keeping pressure constant, the temperature was ramped up to $\simeq$1300~$^o$C in 3~h, held there for a period of 9~h, and then cooled down to the room temperature in 3~h. Afterwards, the pressure was released and the sample removed. On two such synthesized samples the ZF and wTF $\mu$SR experiments under ambient pressure were conducted.

The second type of polycristalline CrAs samples was synthesized by solid state reaction as described in \cite{Saparov_SST_12}. The sample obtained by this method was used in ZF and wTF studies under ambient pressure and in all experimental studies under the pressure.

\vspace{0.5cm}
\noindent {\it Pressure Cell.}
The pressure was generated in a piston-cylinder type of cell made of CuBe alloy, which is especially designed to perform muon-spin rotation experiments under pressure \cite{Andreica01}. As a pressure transmitting medium 7373 Daphne oil was used. The pressure was measured in situ by monitoring the pressure shift of the superconducting transition temperature of In. The maximum safely reachable pressures at $T=300$ and 3~K are 14 and 11~kbar, respectively \cite{Andreica01}.

\vspace{0.5cm}
\noindent {\it Muon-spin rotation ($\mu$SR).}
$\mu$SR measurements at zero field (ZF) and field applied transverse to the initial muon-spin polarization were  performed at the $\pi$M3 and $\mu$E1 beamlines (Paul Scherrer Institute, Villigen, Switzerland), by using the GPS and GPD spectrometers, respectively. At the GPS spectrometer, equipped with a continuous flow $^4$He cryostat, ZF and 3~mT weak transverse field (wTF) experiments at ambient pressure and down to temperatures $1.6$~K  were carried out. At the GPD spectrometer, equipped with an Oxford sorption pumped $^3$He cryostat (base temperature $\sim 0.24$~K) and continuous flow $^4$He cryostat (base temperature $\simeq 2.2$~K), the ZF, 5~mT wTF, and 30~mT transverse field (TF) $\mu$SR experiments under pressure up to $\sim$10.3~kbar were conducted. All ZF and TF experiments were performed by stabilizing the temperature prior to recording the muon-time spectra. In the wTF experiments under pressure the temperature was swept up and down with the rate $\simeq0.2$~K/min. The data were collected continuously. Each muon-time spectra was recorded during approximately 5 minutes.

In a $\mu$SR experiment nearly 100\% spin-polarized muons are implanted into the sample one at a time. The positively charged muons thermalize at interstitial lattice sites, where they act as magnetic microprobes. The muon spin precesses about the local magnetic field $B$ at the muon site with the Larmor frequency $\omega_\mu = \gamma_\mu B$ ($\gamma_\mu/2\pi= 135.5$~MHz/T is the muon gyromagnetic ratio).

In pressure experiments a large fraction of the muons, roughly 50\%, stops in the pressure cell walls adding a background contribution, which has to be separated from the sample signal in the data analysis.
The detailed description of the data analysis procedure is given in the "Supplementary material" part.

 \vspace{1cm}
 \noindent {\bf Acknowledgments}\\
Part of this work was performed at the Swiss Muon Source (S$\mu$S) Paul Scherrer Insitute, Villigen, Switzerland. The work at
the Oak Ridge National Laboratory was supported by the Department of Energy, Office of Science Basic Energy Sciences,
Materials Science and Engineering Division (AS); also partially by the LDRD program (MS).  The authors acknowledge helpful
discussions with L.~Keller, J.~White, and M.~Frontzek.

 \vspace{1cm}
 \noindent {\bf Author contributions}\\ 
 R.K. have performed the experiment, analyzed the data and wrote the paper. Z.G. and H.L. have taken part in $\mu$SR experiments. N.D., M.A.S. and A.S.S. prepared the samples. I.E. have provided the theory description. Z.G., I.E, H.L., A.A., P.K.B, Ch.R, A.S.S, and E.M. took part in discussions and preparation of the manuscript.

\vspace{1cm}
 \noindent {\bf Additional Information}\\ 
Correspondence and requests for materials should be addressed to R.K. (rustem.khasanov@psi.ch).  \\ \\ \\ \\
\\ \\ \\ \\ \\ \\ \\

\setcounter{equation}{0} \setcounter{figure}{0}

\renewcommand{\theequation}{S\arabic{equation}}
\renewcommand{\thefigure}{Sup \arabic{figure}}

\newpage

\begin{center}
{\bf Supplementary materials for "Pressure-induced electronic phase separation of magnetism and superconductivity in CrAs"}
\end{center}

\subsubsection{Muon-spin rotation data analysis procedure}

In pressure experiments a large fraction of the muons, roughly 50\%, stops in the pressure cell walls. The fit function consists, therefore, of the "sample" and the background (pressure cell) contributions and is described as:
\begin{equation}
A(t)=A_s(0) P_s(t)+ A_{pc}(0) P_{pc}(t).
 \label{eq:ZF_Asymmetry_PC}
\end{equation}
Here $A_{s}(0)$ and $A_{pc}(0)$ are the initial asymmetries and $P_{s}(t)$ and $P_{pc}(t)$ are the muon-spin polarizations belonging to the sample and the pressure cell, respectively. The polarization of the pressure cell is generally studied in separated set of experiments.

\subsubsection {ZF $\mu$SR experiments.}
At ambient pressure CrAs is characterized by long-range helimagnetic order with a propagation vector $k_c=0.3562(2)$ parallel to the $c-$axis and the magnetic moments lie in the $ab$ plane \cite{Keller_Arxiv_14_Sup}. The ordered magnetic moment per Cr is $m\simeq1.73$~$\mu_B$ \cite{Keller_Arxiv_14_Sup}. Due to the incommensurability of the magnetic structure, a continuous set of local fields is expected to be seen at each particular muon stopping site. It was shown that such a magnetic structure leads to a field distribution given by \cite{Andreica01_Sup,Schenk_01_sup,Yaouanc_book_11_sup}:
\begin{equation}
P(B)=\frac{2}{\pi}\frac{B}{\sqrt{(B^2-B^2_{min})(B^2_{max}-B^2)}}
 \label{eq:helical_PB}
\end{equation}
and is characterised by two peaks due to the minimum ($B_{min}$) and maximum($B_{max}$) cutoff fields (see the inset in Fig.~1 in the main text). Considering only one muon stopping site, the ZF muon-spin polarization for a powder sample would follow the relation \cite{Yaouanc_book_11_sup}:
\begin{equation}
P_{ZF}(t)=\left[\frac{1}{3} e^{-\lambda_{L}t} +\frac{2}{3} e^{-\lambda_{T}t} \int P(B)\cos\gamma_\mu Bt\right].
 \label{eq:ZF}
\end{equation}
Here $\lambda_T$ and $\lambda_L$ are the transverse and the longitudinal exponential relaxation rates, respectively. The occurrence of 2/3 oscillating and 1/3 non–oscillating $\mu$SR signal fractions originates from the spatial averaging in powder samples, where 2/3 of the magnetic field components are perpendicular to the muon-spin and cause a precession, while the 1/3 longitudinal field components do not.

\begin{figure}[htb]
\includegraphics[width=1\linewidth]{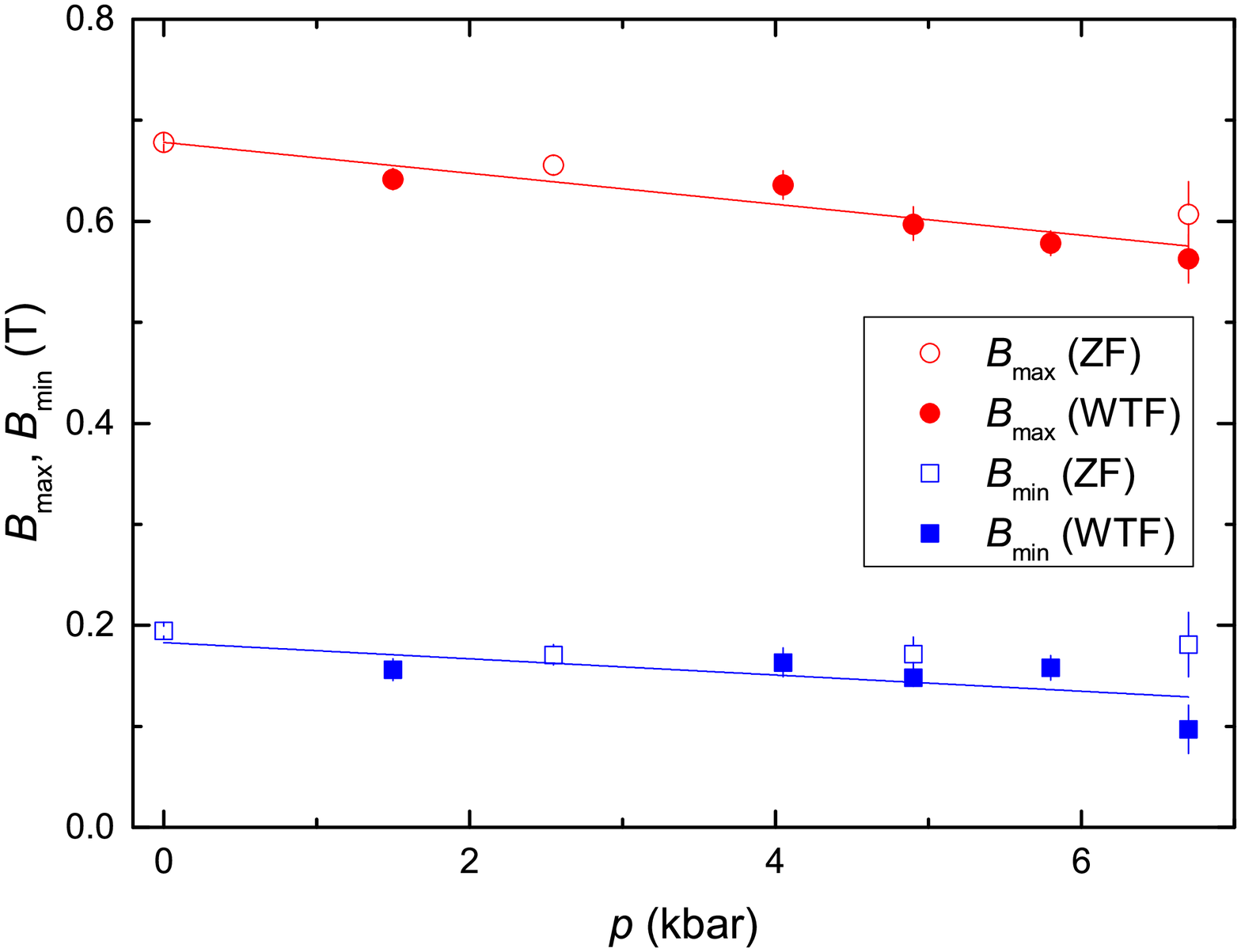}
%
\caption{ Pressure dependence of the minimum $B_{min}$ and the maximum $B_{max}$ cutoff fields of CrAs as obtained from the fit of ZF (open symbols) and wTF (closed symbols) $\mu$SR data for $T\lesssim 5$~K. The solid lines are linear fits with $B_{min}(p)=0.1829(20)-0.0080(8)p$ and $B_{max}(p)=0.6782(14)-0.0153(5)p$. }
 \label{fig:B_min-B_max}
\end{figure}

Figure~\ref{fig:B_min-B_max} shows the dependence of the minimum $B_{min}$ and the maximum $B_{max}$ cutoff fields of CrAs on pressure. Points were obtained from the fit of ZF and wTF $\mu$SR data measured at $T\lesssim 5$~K. Both $B_{min}$ and $B_{max}$ decrease with increasing pressure. Following Ref.~\onlinecite{Brewer_inet_sup} for a helical magnetic structure the upper and the lower cutoff fields should scale as $2m$ and $m$, respectively. Linear fits resulting in ${\rm d}B_{min}/{\rm d}p=-8.0(8)$~mT/kbar and ${\rm d}B_{max}/{\rm d}p=-15.3(5)$~mT/kbar thus confirm this statement.

The decrease of $B_{min}$ and $B_{max}$ with increasing pressure implies a decrease of the ordered magnetic moment. By taking into account that the ambient pressure value of the ordered moment per Cr was found to be $\simeq1.73$~$\mu_B$ \cite{Keller_Arxiv_14_Sup} our results would imply that with increasing pressure up to $p\simeq6.7$~kbar Cr moments decrease down to $\simeq 1.47$~$\mu_B$.

\subsubsection {\it wTF $\mu$SR experiments.}

\begin{figure}[htb]
\includegraphics[width=1\linewidth]{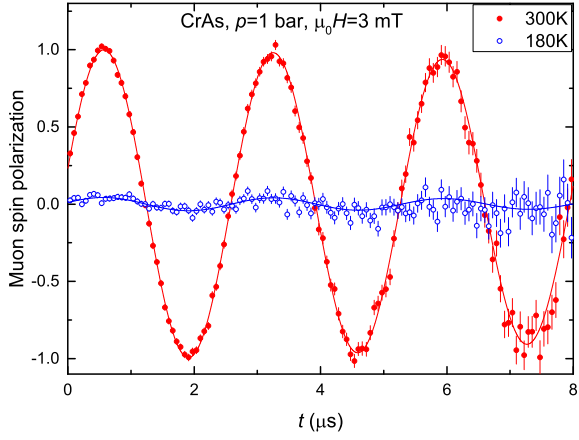}
%
\caption{ WTF $\mu$SR time-spectra ($\mu_0H=3$~mT) of CrAs measured below ($T=180$~K) and above ($T=300$~K) the magnetic transition transition temperature ($T_N\simeq265$~K) at $p=1$~bar. }
 \label{fig:WTF-signal}
\end{figure}

$\mu$SR experiments under weak transverse field (wTF) applied perpendicular to the muon-spin polarization are a straightforward method to determine the onset of the magnetic transition and the magnetic volume fraction. In this case the contribution to the asymmetry from muons experiencing a vanishing internal spontaneous magnetisation can be accurately determined. Muons stopping in a non-magnetic environment produce long lived oscillations, which reflect the coherent muon precession around the external field $B_{ex}$. Muons stopping in magnetically ordered parts of the sample give rise to a more complex, distinguishable signal, reflecting the vector combination of internal and external fields. The random orientation of the grains in a powder sample leads to a broad distribution of precession frequencies.

The situation is substantially simplified for $B_{ex}\ll B_{int}$ (weak transverse field regime). In this case one can neglect the influence of $B_{ex}$ on $B_{int}$ and the fitting function becomes:
\begin{eqnarray}
A(0)P(t)&=&A_{nm}(0)\cos (\gamma_\mu B_{ex}t+\phi)\  e^{-\sigma_{nm}^2t^2/2}.  \nonumber \\
&&+A_{m}(0)P_{ZF}(t)+A_{pc}(0) P_{pc}(t)
 \label{eq:WTF}
\end{eqnarray}
Here $A_{nm}(0)$ and $A_m$ are the initial non-magnetic  and magnetic asymmetry, respectively, $\phi$ is the initial phase of the muon-spin ensemble, and $\sigma_{nm}$ is the temperature independent Gaussian relaxation rate caused by nuclear moments. $P_{ZF}(t)$ represent the ZF magnetic polarization and is described by Eq.~(\ref{eq:ZF}).

Figure~\ref{fig:WTF-signal} represents the wTF $\mu$SR time spectra measured at ambient pressure above ($T\simeq300$~K) and below ($T\simeq180$~K) the magnetic transition ($T_N\simeq265$~K). The solid lines correspond to the fit of the first term on the right-hand side of Eq.~(\ref{eq:WTF}) to the experimental data. The "magnetic term" [$A_{m}(0)P_{ZF}(t)$] vanishes within the first $\sim 0.1$~$\mu$s and thus is not observed with the present data binning ($\simeq 0.063$~$\mu$s). The "pressure cell" contribution is missing since experiments under ambient pressure were performed by using the sample outside of the cell on the low-background GPS spectrometer.

\begin{figure}[htb]
\includegraphics[width=1\linewidth]{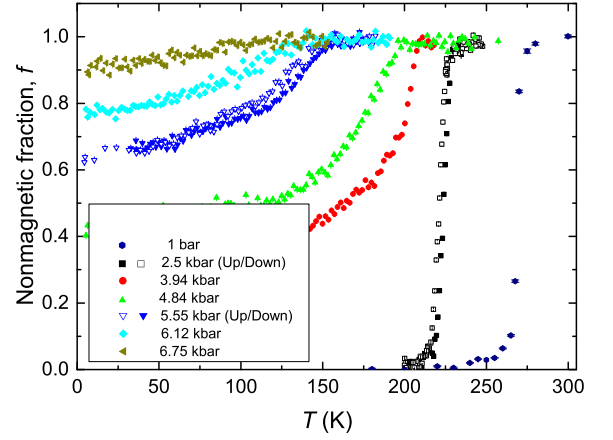}
%
\caption{ Temperature evolution of the non-magnetic volume fraction $f=A_{nm}(0)/[A_{nm}(0)+A_m(0)]$ of CrAs obtained in wTF $\mu$SR measurements at $p=1$~bar, 2.5~kbar, 3.94~kbar, 4.84~kbar, 5.55~kbar, 6.12~kbar, and 6.75~Kbar. Closed and open symbols correspond to the experimental data obtained with increasing and decreasing temperature, respectively. }
 \label{fig:WTF-fraction}
\end{figure}

Figure~\ref{fig:WTF-fraction}  demonstrates the dependence of the non-magnetic volume fraction $f=A_{nm}(0)/[A_{nm}(0)+A_m(0)]$ on temperature at various pressures.

\subsubsection{TF $\mu$SR experiments}

\begin{figure}[htb]
\includegraphics[width=1\linewidth]{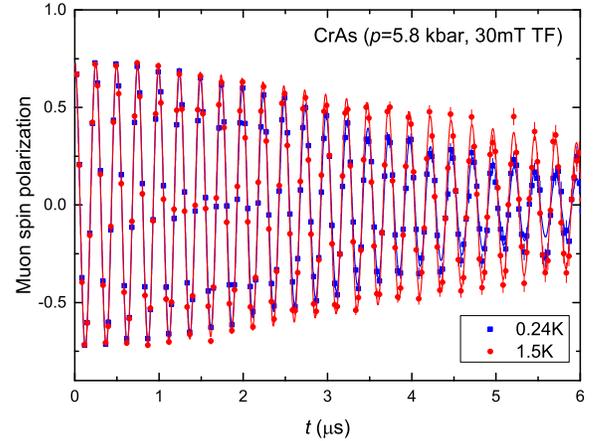}
%
\caption{ TF $\mu$SR time-spectra ($\mu_0H=30$~mT) of CrAs measured below ($T=0.24$~K) and above ($T=1.5$~K) the superconducting transition temperature ($T_c\simeq1.1$~K) at $p=5.8$~kbar. The stronger damping in the superconducting state is due to the formation of the flux line lattice.}
 \label{fig:TF-signal}
\end{figure}

Figure~\ref{fig:TF-signal} shows the TF $\mu$SR time spectra measured at $T=0.24$~K and 1.5~K at $p=5.8$~kbar. The stronger damping at $T=0.24$~K is due to inhomogeneous field distribution caused by formation of the flux line lattice (FLL) in the superconducting CrAs.

The TF $\mu$SR data were analyzed by using the following
functional form:
\begin{eqnarray}
A(0)P(t)&=&A_{nm}(0)\;e^{-[\sigma_{nm}^2+\sigma_{sc}^2]t^2/2}\cos(\gamma
B t+\phi) \nonumber \\
&&+A_{m}(0)P_{ZF}(t)+A_{pc}(0) P_{pc}(t).
 \label{eq:TF}
\end{eqnarray}
Here $A_{nm}(0)$, $A_m$, $\phi$, and $\sigma_{nm}$ have similar meanings as in Eq.~(\ref{eq:WTF}), $\sigma_{sc}$ is the relaxation rate caused by FLL formation, and $B$ is the magnetic field inside the sample. Due to the diamagnetism of the superconducting state $B<B_{ex}$ for $T<T_c$ and $B\simeq B_{ex}$ for $T\geq T_c$.


\begin{thebibliography}{99}
\bibitem{comment1} There is approximately 0.8~kbar systematic shift between $\chi_{dia}(p)$ reported in
    Ref.~\onlinecite{Wu_Arxiv_14} and $f(p)$ measured here. It could be caused by the sample difference or/and by the
    pressure determination procedure.
\end{thebibliography}
\end{document}